\documentclass{elsart}
\usepackage{natbib}
\begin{document}
\begin{flushright}
	FERMILAB-Conf-00/179-T\\[0.2in]
\end{flushright}
\begin{frontmatter}
\title{Overview of Grand Unified Models and\\
	Their Predictions for Neutrino Oscillations\thanksref{workshop}}
\author{Carl H. Albright\thanksref{e-mail}}
\address{Fermilab, P.O. Box 500, Batavia, IL 60510}
\thanks[workshop]{Contribution to the Proceedings of the NuFACT'00 
	Workshop held in Monterey,\\ \hspace*{0.1in} CA May 22-26, 2000}

\thanks[e-mail]{e-mail: albright@fnal.gov}

\begin{abstract}
A brief overview of Grand Unified Models is presented with some attention
paid to their predictions for neutrino oscillations.  Given the well-known
features of the two non-unified standard models, SM and MSSM, a 
listing of the features of classes of unified models is given, where a GUT 
flavor symmetry and/or family symmetry are introduced to reduce the number 
of model parameters.  Some general remarks are then made concerning the 
type of predictions that follow for the neutrino masses and mixings.
\end{abstract}

\end{frontmatter}
\vspace*{-0.5in}
\section{Non-unified Standard Models}
\vspace*{-0.3in}
We begin with some relevant features of the two non-unified standard
models.  In the SM of particle physics, where 
the gauge group is $SU(3)_c \times SU(2)_L \times U(1)_Y$, no gauge 
coupling unification occurs at any scale [1].  Just one Higgs doublet 
of $SU(2)_L$ is introduced, $H_u = \phi = (\phi^+,\phi^0)^T$, along with 
its charge conjugate, $H_d = \tilde{\phi} = (\phi^0,-\phi^-)^T$, with the 
symmetry broken by the vacuum expectation value (VEV),
$\langle \phi^o \rangle = v = 174$ GeV.  The left-handed quarks and 
leptons are placed in doublets, 
while the right-handed (or left-handed conjugate) quarks and charged 
leptons are taken to be singlets.  The Yukawa Lagrangian can then be 
written as 

\begin{equation}
  \mathcal{L}_Y = \lambda^u_{ij} u^c_i Q_j H_u + \lambda^d_{ij} d^c_i Q_j H_d
		+ \lambda^e_{ij} d^c_i L_j H_d
  \label{e1}
\end{equation}

\noindent
where $Q_j$ and $L_j$ refer to the jth family of quark and lepton 
left-handed doublets.
Since no right-handed (or left-handed conjugate) neutrinos are 
assumed to exist, no renormalizable neutrino mass terms appear in eq. 
(\ref{e1}).

In the case of the minimal supersymmetric standard model (MSSM), the same
gauge group applies while two independent Higgs doublets, $H_u$ and $H_d$ 
are introduced with VEV's, 
$v_u = \langle H^o_u \rangle,\ v_d = \langle H^o_d \rangle$,
which give masses to the up-type quarks and down-type quarks, respectively.
The constraint $v = \sqrt{v^2_u + v^2_d} = 174\ {\rm GeV}$ holds, while
the parameter $\tan \beta = v_u/v_d$ remains arbitrary.  The quark,
lepton, scalar and gauge boson sectors are also doubled by the introduction
of superpartners.  With these extra SUSY particles present, gauge coupling
unification occurs at a scale of $\Lambda_G = 2 \times 10^{16}$ GeV 
[1].  The Yukawa superpotential has the same form as for the SM, 
still with many arbitrary Yukawa couplings and no renormalizable neutrino 
mass terms.
\\[-0.5in]
\section{Reduction in the Number of Parameters}
\vspace*{-0.3in}
In order to reduce the number of free parameters in the standard models, 
one can introduce a flavor (or intra-family) symmetry, a family (or
inter-family) symmetry, or both.  Such procedures have mainly 
been carried out in the supersymmetry framework, since that allows the 
possibility of gauge coupling unification and can sufficiently suppress
proton decay.  

Flavor symmetry has generally been achieved in the framework of Grand 
Unified Theories (GUTs) which provide unified treatments of quarks
and leptons, as (some) quarks and leptons are placed in the same multiplets.
Examples involve $SU(5),\ SU(5) \times U(1),\ SO(10),\ E_6,\ SU(5) \times 
SU(5)$, {\it etc}.  

The introduction of a family or horizontal symmetry, on the other hand,
enables one to build in an apparent hierarchy for different family masses
belonging to comparable flavors.  Such a symmetry may be discrete
as in the case of $Z_2,\ S_3,\ Z_2 \times Z_2$, {\it etc.} which results in 
multiplicative quantum numbers.  A continuous symmetry such as 
$U(1),\ U(2),\ SU(3)$, etc., on the other hand, results in additive 
quantum numbers and may be global or local (and possibly anomalous).  

Combined flavor and family symmetries will typically reduce the number of
model parameters even more effectively.  On the other hand, the unification
of flavor and family symmetries into one single group such as $SO(18)$ or
$SU(8)$, for example, has generally not been successful, as too many extra 
states are present which must be made superheavy.

We now turn to illustrate various features of the different types of 
unification.  For lack of space only a few selected references are given.  
A considerably more complete set of references can be found in a more 
general recent review [2].
\\[-0.5in]
\section{MSSM with Anomalous U(1) Family Symmetry}
\vspace*{-0.3in}
In 1979 Froggatt and Nielsen [3] added to the SM a scalar singlet 
``flavon'' $\phi_f$, which gets a VEV, together with heavy fermions, 
$(F,\ \bar{F})$, in vector-like representations, all of which carry 
$U(1)$ family charges.  With $U(1)$ broken at a scale $M_G$ by 
$\langle \phi \rangle/M_G \equiv \epsilon 
\sim (0.01 - 0.1)$, the light and heavy fermions are mixed; hence
$\epsilon$ can serve as an expansion parameter for the quark and 
lepton mass matrix entries.

This idea received a revival in the past decade when it was observed by 
Ibanez [4] that string theories with anomalous 
$U(1)$'s generate 
Fayet-Iliopoulos D-terms which trigger the breaking of the $U(1)$ at a scale 
of $O(\epsilon)$ below the cutoff, again providing a suitable expansion 
parameter.  The $\epsilon^n$ structure of the mass matrices can be determined 
from the corresponding Wolfenstein $\lambda$ structure of the 
CKM matrix and the quark and lepton mass ratios, where different
$U(1)$ charges are assigned to each quark and lepton field.
But this procedure suffers from the fact that the coefficients (prefactors) 
of the $\epsilon$ powers are not accurately determined.

With this scenario, Ramond [5] and many others have shown that 
maximal $\nu_\mu \rightarrow \nu_\tau$ mixing of atmospheric neutrinos and the 
small mixing angle MSW solution for solar $\nu_e \rightarrow \nu_\mu,\ 
\nu_\tau$ can be obtained.  Bimaximal mixing solutions, however, are not 
obtained naturally.
\\[-0.5in]
\section{Minimal SUSY SU(5)}
\vspace*{-0.3in}
With the minimal SUSY $SU(5)$ flavor symmetry, the matter fields are placed
in ${\bf \bar{5}}$ and ${\bf 10}$ representations: $\bar{5}_i \supset 
(d^c_\alpha,\ \ell,\ \nu_\ell)_i,\quad 10_i \supset (u_\alpha,\ d_\alpha,
\ u^c_\alpha,\ \ell^c)_i$.  Higgs fields are placed in the adjoint and 
fundamental representations: $\Sigma(24),\ H_u(5)$, $H_d(\bar{5})$.  The 
$SU(5)$ symmetry is broken down to the MSSM at a scale $\Lambda_G$
with $\langle \Sigma \rangle = (b,b,b,-\frac{3b}{2},-\frac{3b}{2})$, 
but doublet-triplet splitting must be done by hand.  The 
electroweak breaking occurs when $H_u$ and $H_d$ VEV's are generated.

The number of Yukawa couplings has now been reduced in the Yukawa 
superpotential

\begin{equation}
	W_{Yuk} = \lambda^u_{ij} 10_i \cdot 10_j \cdot H_u + \lambda^d_{ij}
		\bar{5}_i \cdot 10_j \cdot H_d
	\label{su5 yuk}
\end{equation}

\noindent
where by the particle assignments, the fermion mass matrices 
exhibit the symmetries, $M_U = M^T_U,\ M_D = M^T_L$ which implies at the 
GUT scale $m_b = m_\tau\ {\rm but\ also}\ m_d/m_s = m_e/m_\mu$, 
since minimal $SU(5)$ is too simplistic and no family symmetry is present.

Proton decay occurs through leptoquark gauge boson exchange with $\tau_p = 
10^{36 \pm 1.5}$ yrs. and through colored Higgsino exchange with $\tau_p = 
10^{32-33}$ yrs. [6]  This places it nearly within reach but well 
above the old non-SUSY $SU(5)$ limit of $10^{29}$ years which has long been 
ruled out.
\\[-0.5in]
\section{SUSY SU(5) with Anomalous U(1)}
\vspace*{-0.3in}
Kaplan, Lepintre, Masiero, Nelson and Riotto [7] have suggested an 
extension of the Froggatt-Nielsen idea as applied to $SU(5)$ by addition of 
more singlet Higgs flavons with different VEV's to replace different powers 
of the same ratio.  The effective couplings then appear as 
$f^c f H{\langle \phi \rangle}/{\langle\chi \rangle}$, where $W_{Yuk} = 
f^c F H + F^c F \chi + F^c f \chi$.
Matter and Higgs supermultiplets are each assigned their own family $U(1)$ 
charges.  Although there are more restrictions,
the model is still not very predictive due to the unknown prefactors.
\\[-0.5in]
\section{Flipped SU(5) $\times$ U(1) with Anomalous U(1)}
\vspace*{-0.3in}
This partially unified $SU(5) \times U(1)$ GUT [8] has the following 
unconventional assignments:
Matter Fields: $\bar{5}_i \supset (u^c_\alpha,\ \nu_\ell,\ \ell)_i$;
$10_i \supset (u_\alpha,\ d_\alpha,\ d^c_\alpha,\ \nu^c)_i$; 
$1_i \supset (\ell^c)_i$;
Higgs Fields: $\Sigma(10),\ \bar{\Sigma}(\overline{10}),\quad H(5),\quad 
\bar{H}(\bar{5})$.
No $b - \tau$ unification or seesaw mechanism occurs, while R-parity is broken.
\\[-0.5in]
\section{SUSY SO(10) with Family Symmetry}
\vspace*{-0.3in}
Here all fermions of one family are placed in a ${\bf 16}$ spinor 
supermultiplet and carry the same family charge assignment:
${\bf 16}_i(u_\alpha,\ d_\alpha,\ u^c_\alpha,\ d^c_\alpha,\ \ell,
\ \ell^c,\ \nu_\ell,\ \nu^c_\ell)_i$, $i = 1,2,3.$
Massive pairs of $({\bf 16},\ \overline{\bf 16})$'s and ${\bf 10}$'s 
may also be present. The Higgs Fields:
${\bf 45}_H$'s and ${\bf 16}_H,\ \overline{\bf 16}_H$ break $SO(10)$ to SM,
while $\bf{10}_H$ breaks the electroweak group.
A $\overline{\bf 126}_H$ or $\overline{\bf 16}_H\cdot
\overline{\bf 16}_H\cdot 1_H$ can generate superheavy right-handed
Majorana neutrino masses.

$t-b-\tau$ Yukawa coupling unification is possible only for 
$\tan \beta = v_u/v_d \simeq 55$ in this minimal case.  However, if a 
${\bf 16'}_H,\ \overline{\bf 16'}_H$ pair is introduced with the former
getting an electroweak-breaking VEV which helps contribute to $H_d$ [9],
Yukawa coupling unification is possible for $\tan \beta \ll 55$.
Such breaking VEV's can contribute asymmetrically 
to the down quark and charged lepton mass matrices.  This makes it 
possible to understand large $\nu_\mu - \nu_\tau$ mixing, $U_{\mu 3} 
\simeq 0.707$, while $V_{cb} \simeq 0.040$.
Moreover, the Georgi-Jarlskog mass relations [10],
$m_s/m_b = m_\mu/3m_\tau\ {\rm and}\ m_d/m_b = 3m_e/m_\tau$,
can be generated by the mass matrices with the help of the asymmetrical
contributions just mentioned.

Various family symmetries appear in the literature in combination with 
the $SO(10)$ group: unspecified [11], 
$U(1) \times Z_2 \times Z_2$ [9], 
$U(2)$ [12],
$U(2) \times U(1)^n$ [13],
$SU(3)$ [14], to name just a few.  Again see [2].
\\[-0.5in]
\section{Mass and Mixings Matrices}
\vspace*{-0.3in}
Complex symmetric mass matrices can be constructed for each flavor 
$f = U,\ D,\ N,\ L$ in the GUT flavor basis 

\begin{equation}
	{\mathcal B}_f = \left\{f_{iL},\ f^c_{iL}\right\},\ i = 1,2,3: \qquad
	{\mathcal M}_f = \left(\matrix{0 & M^T_f\cr M_f & M_{R}\cr}\right)
	\label{mass}
\end{equation}

\noindent
where $M_f$ is the $3 \times 3$ Dirac mass matrix for flavor $f$ and $M_{R}$ 
is the $3 \times 3$ superheavy Majorana neutrino mass matrix for $f = N$.
The $3 \times 3$ light neutrino mass matrix is determined by the seesaw 
mechanism, $M_\nu = - M^T_N M^{-1}_R M_N$.

The mixing matrices are obtained by diagonalizing the mass matrices according 
to $M^{diag}_f = U^T_f M_f U_f$ from which one finds
$V_{CKM} = U^\dagger_U U_D,\qquad U_{MNS} = U^\dagger_L U_\nu$.
Successful GUT models must essentially generate
the CKM mixing matrix and one of the two MNS maximal (SMA MSW) or 
bimaximal (LMA MSW or Vacuum) mixing matrices corresponding to 
the still possible solar neutrino solutions:\\[-0.3in]

\begin{eqnarray}
	V_{CKM} &\simeq& \left(\matrix{0.975 & 0.220 & 0.0032e^{-i65^o}\cr
		-0.220 & 0.974 & 0.040\cr 0.0088 & -0.040 & 0.999\cr}\right)\\
	U^{(SMA)}_{MNS} &\simeq& \left(\matrix{0.99 & 0.04 & 0.05\cr
		-0.03 & 0.70 & -0.71\cr -0.03 & 0.71 & 0.70\cr}\right)\\
	U^{(bimax)}_{MNS} &\simeq& \left(\matrix{0.71 & -0.70 & \theta_{13}\cr
		0.50 & 0.50 & -0.71\cr 0.50 & 0.50 & 0.70\cr}\right)
\end{eqnarray}

\noindent
where $\theta_{13}$ is bounded to be $< 0.2$ by the CHOOZ experiment
[15].
The effective mixing angles and mass squared differences in question 
are:\\[-0.3in]

\begin{eqnarray}
	\sin^2 2\theta_{\rm atm} &\equiv& 4|U_{\mu 3}|^2 |U_{\tau 3}|^2 
	 	\simeq 1.0\\
	\sin^2 2\theta_{\rm sol} &\equiv& 4|U_{e1}|^2 |U_{e2}|^2 
		\simeq \left\{\matrix{0.006, & {\rm SMA} \qquad\cr
		1.0, & {\rm LMA,\ Vac}\cr}\right.\\
	\sin^2 2\theta_{\rm reac} &\equiv& 4|U_{e3}|^2 (1 - |U_{e3}|^2)
		\sim  \left\{\matrix{0.05, & {\rm SMA\qquad}
		\cr
		0.001 - 0.1, & {\rm LMA,\ Vac}\cr}\right.\\
	\Delta m^2_{32} &\simeq& 3 \times 10^{-3}
		\ {\rm eV^2},\quad \Delta m^2_{21} \simeq \left\{\matrix{6 
		\times 10^{-6}\ {\rm eV^2}, & {\rm SMA}\cr
		5 \times 10^{-5}\ {\rm eV^2}, & {\rm LMA}\cr
		4 \times 10^{-10}\ {\rm eV^2}, & {\rm Vac}\cr}\right.
\end{eqnarray}

The SMA MSW solution is readily found in many unified models, if the 
light neutrino diagonalizing matrix, $U_\nu$, is close to the identity matrix 
and the maximal atmospheric neutrino mixing is provided by the charged lepton
diagonalizing matrix.  Bimaximal mixing can be obtained with substantial
off-diagonal contributions in the $M_{NR}$ matrix.  Of the two, the Vacuum 
solution tends to be somewhat more natural than the LMA MSW
solution, although either can be obtained with careful tuning of the Dirac 
and Majorana matrices.
\\[-0.5in]
\section{Partial Unification with Light Sterile Neutrinos}
\vspace*{-0.3in}
Light sterile neutrinos are somewhat of an anomaly in the presence of GUTs.  
They tend to spoil the seesaw mechanism, so it is not clear why the ordinary 
and sterile neutrinos remain light.  Moreover, if one sterile neutrino is
found, one expects many.

In $SO(10)$, light sterile neutrinos can be placed into ${\bf 1}$ 
representations, but the number is unexplained.  In $E_6$ one can 
associate one sterile neutrino with each ${16}$ of $SO(10)$ in the ${\bf 27}$,
since ${\bf 27} \supset 16 + 10 + 1$ for each family [16].  But one must 
insure that the members of the $10$ of each ${\bf 27}$ become supermassive.
Other groups suggested in the literature involve $SU(5) \times SU(5)$
and $SO(10) \times SO(10)$, where the sterile neutrinos live in the second
``mirror'' groups and remain light while all other particles in the 
representations decouple from the known interactions [17].
\\[-0.5in]
\section{Concluding Remarks}
\vspace*{-0.3in}
In order to obtain predictive models of the 12 ``light'' fermions
and their 8 CKM and MNS mixing angles and phases in the framework
of this talk, a Grand Unified Model with family symmetry must be introduced.
$SO(10)$ models are more tightly constrained than $SU(5)$ models
and are more economical than larger groups like $E_6$, where more
fields must be made supermassive.  In fact, some $SO(10)$ models 
do very well in predicting the 20 ``observables'' with just 10 or 
more input parameters [18].\\[-0.5in]

\end{document}